\documentclass[preprint2]{aastex}

\newcommand{\swift}{{\em Swift}}
\newcommand{\chandra}{{\em Chandra}}
\newcommand{\INT}{{\it INTEGRAL}}
\newcommand{\asca}{{\it ASCA}}
\newcommand{\einst}{{\it Einstein}}
\newcommand{\fe}{{\it Fermi}}
\def\ergs{erg\,s$^{-1}$}
\def\ergscm2{erg\,s$^{-1}$cm$^{-2}$}

\newcommand{\sgr}{\mbox{SGR\,J$1550-5418$}}
\newcommand{\psr}{\mbox{PSR\,J$1550-5418$}}
\newcommand{\ax}{\mbox{AX\,J$155052-5418$}}
\newcommand{\esrc}{\mbox{1E\,$1547.0-5408$}}

\newcommand{\degrees}{\ensuremath{^\circ}}
\newcommand{\myemail}{azk@mpe.mpg.de}
\usepackage{arydshln}

\shorttitle{SGR J$1550-5418$ bursts with Fermi/GBM}
\shortauthors{von Kienlin et al.}

\begin{document}

\title{Detection of spectral evolution in the bursts emitted during the 2008-2009 active episode of SGR~J$1550-5418$}

\author{Andreas von Kienlin\altaffilmark{1},
David Gruber\altaffilmark{1},
Chryssa Kouveliotou\altaffilmark{2},
Jonathan Granot\altaffilmark{3},
Matthew G. Baring\altaffilmark{4},
Ersin G\"o\u{g}\"u\c{s}\altaffilmark{5},
Daniela Huppenkothen\altaffilmark{6},
Yuki Kaneko\altaffilmark{5},
Lin Lin\altaffilmark{5},
Anna L.~Watts\altaffilmark{6},
P.~Narayana Bhat\altaffilmark{7},
Sylvain Guiriec\altaffilmark{7,8},
Alexander J.~van der Horst\altaffilmark{6,9},
Elisabetta Bissaldi\altaffilmark{1,10},
Jochen Greiner\altaffilmark{1},
Charles A. Meegan\altaffilmark{9},
William S.~Paciesas\altaffilmark{7},
Robert D.~Preece\altaffilmark{7},
Arne Rau\altaffilmark{1}}
\email{\myemail}
\altaffiltext{1}{Max-Planck-Institut f\"ur extraterrestrische Physik, Giessenbachstra{\ss}e,  85748 Garching, Germany}
\altaffiltext{2}{Space Science Office, VP62, NASA/Marshall Space Flight Center, Huntsville, AL 35812, USA}
\altaffiltext{3}{The Open University of Israel, 1 University Road, POB 808, Ra'anana 43537, Israel}
\altaffiltext{4}{Department of Physics and Astronomy, Rice University, MS-108, P.O. Box 1892, Houston, TX 77251, USA}
\altaffiltext{5}{Sabanc\i University, Faculty of Engineering and Natural Sciences, Orhanl\i$-$Tuzla, \.{I}stanbul 34956, Turkey}
\altaffiltext{6}{Astronomical Institute "Anton Pannekoek," University of Amsterdam, Postbus 94249, 1090 GE Amsterdam, The Netherlands}
\altaffiltext{7}{Center for Space Plasma and Aeronomic Research, University of Alabama in Huntsville, 320 Sparkman Drive, Huntsville, AL 35805, USA}
\altaffiltext{8}{NASA Goddard Space Flight Center, Greenbelt, MD 20771, USA}
\altaffiltext{9}{Universities Space Research Association, NSSTC, 320 Sparkman Drive, Huntsville, AL 35805, USA}
\altaffiltext{10}{Institute of Astro and Particle Physics, University Innsbruck, Technikerstrasse 25, 6176 Innsbruck, Austria}

\begin{abstract}

In early October 2008, the Soft Gamma Repeater \sgr~(\esrc, \ax, \psr) became active, emitting a series of bursts
which triggered the \fe~Gamma-ray Burst Monitor (GBM) after which a second especially intense activity period commenced in 2009 January  and a third, less active period was
detected in  2009 March-April. Here we analyze the GBM data of all the bursts from the first and last active episodes. We performed temporal and spectral analysis for all events and found that their temporal characteristics are very similar to the ones of other SGR bursts, as well  the ones reported for the bursts of the main episode (average burst durations  $\sim170$\,ms). In addition, we used our sample of bursts to quantify the systematic uncertainties of the GBM location algorithm for soft gamma-ray transients to $\lesssim 8\degrees$.
Our spectral analysis indicates significant spectral evolution between the first and last set of events. Although the 2008 October events are best fit with a single blackbody function, for the 2009 bursts an Optically Thin Thermal Bremsstrahlung (OTTB) is clearly preferred. We attribute this evolution to changes in the magnetic field topology of the source, possibly due to effects following the very energetic main bursting episode.

\end{abstract}


\section{Introduction}

Soft Gamma Repeaters (SGRs) together with Anomalous X-ray Pulsars (AXPs) comprise a small group of X-ray pulsars with many observational similarities.  They have slow spin periods clustered in a narrow range ($P\sim$ $2-12$\,s), and relatively large period derivatives ($\dot P \sim 10^{-13}-10^{-10}$\,s\,s$^{-1}$). Their inferred surface dipole magnetic fields of $10^{14}-10^{15}$\,G (\citealt{kou98,kou99}), place these sources at the extreme end of the distribution of magnetic fields in astrophysical objects. Such objects were predicted theoretically by \citet{duncan92}, \citet{bohdan92}, and by \citet{usov92}; the former team named such high $B-$field sources ``magnetars''. Phenomenologically, a distinct difference between AXPs and SGRs is manifested by their bursting activity behavior. SGRs have been observed to undergo active episodes with a multitude of short, soft bursts with energies upwards $10^{36}$ erg, reaching over $10^{45}$ erg in the case of the very rare Giant Flares. AXPs, on the other hand, are not such prolific and energetic bursters. As a result, SGRs were historically discovered from their burst activity, while AXPs were set aside from rotation-powered X-ray pulsars by their timing properties with no bursts. This changed in 2002, when \cite{gavriil02} discovered bursts from the AXP\,1E1048.1$-$5937; today almost all AXPs have also been shown to emit SGR-like bursts, albeit fainter. As a group, magnetars are persistent X-ray emitters with X-ray luminosities ranging between $10^{32}-10^{36}$\,\ergs, larger than those obtained by rotational energy losses, supporting the hypothesis that magnetic field dissipation powers their X-ray emission. Alternative models, such as e.g., accretion from a fossil supernova fall-back disk, have also been invoked to explain the magnetar emission; for comprehensive reviews see \cite{woo06} and \cite{mer08} and references therein.

\sgr~is a source that has undergone multiple identity changes. The source was discovered in 1980 with the \einst~observatory (\einst~source: \esrc) during a search for X-ray counterparts of {\it COS-B} unidentified $\gamma$-ray sources \citep{lam81}. The Advanced Satellite for Cosmology and Astrophysics (\asca) confirmed the detection during a Galactic Plane survey in 1998 (\asca~source: AX J155052-5418, \citealt{sug01}). \cite{gel07} proposed, on the basis of {\it XMM-Newton} and {\it Chandra} X-ray Observatory ({\it CXO}) observations in 2004 and 2006, a potential magnetar/SNR association between this X-ray source and the Galactic radio shell G$327.24-0.13$. The location in the center of the SNR candidate, the relatively soft X-ray spectrum and the X-ray variability favored a magnetar model explanation, however the {\it XMM} observation only set an upper-limit for the peak-to-peak pulsed fraction. The crucial piece of evidence was finally provided by radio observations  (PSR\,J$1550 - 5418$; \citealt{cam07}). Their measurement of the spin period of 2.07 s and  period derivative of $2.3 \times 10^{-11}$ s s$^{-1}$  led to an estimate for the surface magnetic dipole  field of $B \sim 2.2 \times 10^{14}$~G, which confirmed the magnetar nature of the source.

Due to the lack of bursting activity, the source was initially characterized as an AXP \citep{cam07}. However, in 2008/2009 the source entered an extremely active period emitting a plethora of SGR-like bursts, in three active episodes, of  which the first and the last were the least prolific. Based on the main episode behavior, during which several hundreds of bursts were emitted during a 24-hour period, similar to the burst ``storms'' of SGRs $1627-41$ \citep{esposito08}, SGR $1900+14$ \citep{gogus99,israel08} and SGR $1806-20$ \citep{gogus00}, the source was renamed as \sgr~(\citealt{pal09}, \citealt{kou09}). A detailed study of the 2008 October \swift~X-ray bursts from SGR\,J$1550-5418$ is presented in \cite{isr10}. A detailed analysis of $\sim200$ bursts detected within a few hours on January 22 with the International Gamma-Ray Astrophysics Laboratory (\INT), is presented in \cite{mer09} and \cite{sav10}. Finally, \cite{ter09} present the analysis of over 250 bursts  (50 keV to 5 MeV) detected with the {\it Suzaku}/Wide-band All-sky Monitor (WAM), while the observations of the \swift/BAT, Konus-Wind, and {\it RHESSI} are published by \cite{gro09}, \cite{gol09}, and \cite{bel09}, respectively.

\chandra~and {\it RXTE} observations, performed after the 2008 and 2009 outbursts, provided evidence of a decoupling between magnetar spin and radiative properties \citep{ng11}. The pulsar spin-down was observed to increase by a factor of 2.2 during the first 2008 event, in absence of a corresponding spectral change.  During the more energetic 2009 events no such variation was found.  A comprehensive analysis of multiple instrument X-ray data by \cite{ber11} recorded since the 1980 discovery of the source, showed that the X-ray flux history can be grouped into three levels: low, intermediate and high. The observed persistent spectra harden when transitioning from the low to the high state, with the power-law component becoming flatter and the temperature of the blackbody increasing. During the high flux state the pulsed fraction decreases with energy and shows an anti-correlation with the X-ray flux.

The Gamma-ray Burst Monitor (GBM) onboard {\it Fermi} detected all three active episodes of the source. A detailed temporal and (integrated) spectral study of the main (second) burst episode (2009 January 22 to 29) comprising 286 bursts was published by \cite{AvdH12}. A search in the \fe~/LAT data in the $0.1-10$ GeV energy range did not reveal significant gamma-ray emission \citep{abd09}. During this episode, the first GBM trigger on 2009 January 22, showed a $\sim 150$~s long persistent emission with intriguing timing and spectral properties. \cite{kan10} identified coherent pulsations up to $\sim 110$~keV at the spin period of the neutron star and an additional (to a power-law) blackbody component required to model the enhanced emission spectra. The favored  emission scenario is a  surface hot spot with the dimensions of the magnetically confined plasma  near the neutron star surface (roughly a few $\times 10^{-5}$ of the neutron star area).  \cite{tie10} claimed two best fit source distances of 3.9~kpc and $\sim 5$~kpc, using the X-ray rings around \sgr~ produced by dust-scattering. Here we adopt a distance of 5~kpc.

Such an extensive burst activity with hundreds of bursts within a few days has only been observed from three more sources in the past: SGRs\,$1806-20, 1900+14$ and $1627-41$. However, lower level burst activity prior to the main bursting episode was only seen from SGRs\,$1900+14$ and $1806-20$. Detailed spectral analyses of the bursts from these sources showed that the spectral properties of the earlier bursts were not distinct from those occurring during their peak activity episodes   (G{\"o}{\u g}{\"u}{\c s} et al. 1999, 2000). In contrast, in SGR\,J$1550-5418$ the ``before'' and ``after'' burst episodes (2008 October  and 2009 March -- April ) were significantly different than the main ``storm'' of activity: the burst rates and event intensities were much lower. We report here our spectral analysis results of the first and last bursting episode from \sgr, where we see for the first time a clear distinction between the spectral shapes of events detected before and after the main active episode of the source. Therefore, we find here an intriguing new type of behavior that was not observed before in other SGRs. In \S\ref{sec:obs} we present an overview of the source burst activity as seen with GBM, while \S\ref{sec:analysis} presents detailed temporal and spectral analysis of the bursts. Since this is the first soft transient source detected with GBM, we discuss here also the GBM location accuracy for soft sources. We discuss our results and in particular the implications of the spectral differences between the two episodes in \S\ref{sec:discussion}.

\section{Overview of the GBM Observations of \sgr~bursts}\label{sec:obs}
On 2008 October 3, GBM triggered four times on soft bursts consistent with the same Galactic location \citep{vKiBri08}. The source triggered also the {\it Swift} satellite \citep{kri08a,kri08b} approximately 25 min after the first GBM trigger, enabling an accurate location and the subsequent identification of the source with the magnetar candidate \esrc~ \citep{rea08}.
The GBM locations were all consistent with the \esrc~position (see section \ref{sec:burstlocations}).
None of the GBM bursts on 2008 October were detected with \swift, as the two spacecrafts were on opposite sides of the Earth; an untriggered burst search did not reveal any additional bursts. The last GBM burst activity was recorded on 2008 Oct 10, 12:53:38 UT \citep{AvdH08}.

\begin{figure}[t]
\plotone{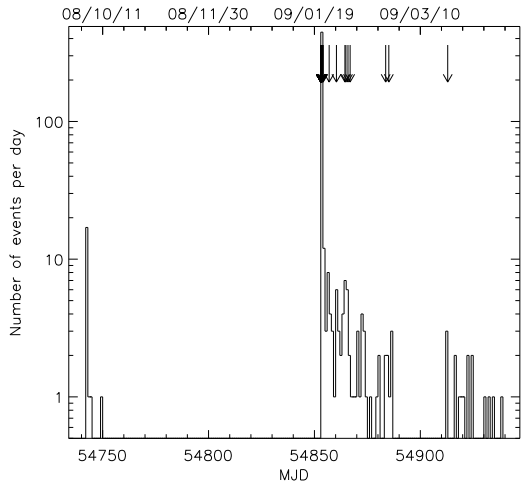}
\caption{Number of \sgr~bursts per day over $\sim210$ days.  Three active periods are clearly visible, with an activity peak at the beginning of the second period (MJD 54850). Arrows indicate times of 61 exceptionally bright events (note that there is significant arrow overlaps in the figure). \label{fig1}}
\end{figure}

The source became active again on 2009 January 22, in a second period of extremely high bursting activity (\citealt{ConBri09, vKiCon09, kou09, mer09}), with several short and very bright bursts occurring at intervals of a few seconds to about 150~s between average bursts. In the next $\sim30$~d,  the instrument triggered on 117 discrete bursts through February 24, of which 15 were extremely intense, saturating the data rate. During the first 24~hr (January 22) only 41 triggers were recorded since GBM flight software requires a minimum of 596~s between triggers.  A search for untriggered events revealed a total of $\sim$450 bursts during this 24~hr period (\citealt{kan10,AvdH12}). The source remained active until 2009 February 24, albeit with a decreased rate of roughly two triggers per day, with exceptionally intense bursts occurring about once per week .

After a month of quiescence the source entered a third period of medium activity, starting on 2009 March 22, 18:56:23.75 UT and lasting until 2009 April 17. During this period the GBM triggered 14 times with  $\sim 4$~hr to $\sim 6$~d between individual triggers. The search for untriggered outbursts did not reveal any additional event. Only one bright burst caused a saturation of the onboard science data bus. During this last period, the mean burst fluence and peak flux were brighter, compared to the October 2008 activity, but fainter, when compared to the 2009 January -- February period. The three periods are shown in Figure~\ref{fig1} as the number of events per day, exhibiting a distinct peak at the beginning of the second activity period (MJD 54850).

Table~\ref{tbl-1} shows the list of GBM SGR\,J$1550-5418$ triggers and bursts during the 2008 October and 2009 March -- April activity periods. The first trigger on 2008 October 3, includes 15 individual bursts. None of these bursts showed emission in the  two GBM BGO detectors ($0.2-40$\,MeV), so the data from these detectors were excluded from our spectral analysis.

The search for untriggered events was performed from 2008 September 22 to October 13 and from 2009 January 15 to April 22 (2008 October 06 and 2009 March 12 and 13 were left out due to data problems), thus covering part of the first gap and the full second gap of bursting activity,  visible in  Figure~\ref{fig1}.
For the 93~d, from 2008 October 14 until 2009 January 14, left out by the untriggered burst search, we can only exclude the occurrence of additional events by the fact that GBM and any other instrument did not trigger on bursts from this source. This is corroborated by the finding that all burst detected by the untriggered bursts search during the activity periods presented here, did trigger GBM or occurred during the $\sim 600$~s time period where GBM  switched into trigger mode.

\begin{figure}[t]
\plotone{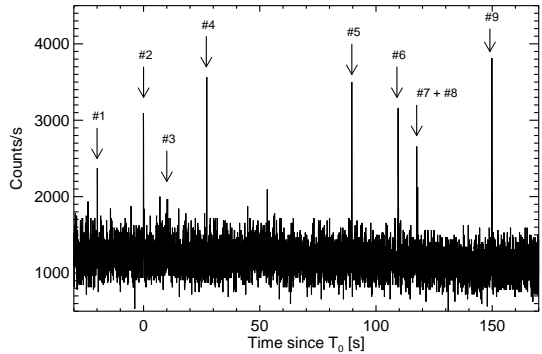}
\caption{The first GBM trigger (bn081003.377) from \sgr, on 2008 October 3, at 09:03:06 UT (244717387 MET). Besides the triggering event (\#2), eight additional, untriggered, bursts (\#1, \#3 - \#9) are also visible (marked with arrows), including burst \#1 observed prior to the trigger time ($T_0 \sim 0$~s),
at $ -19.9$~s. \label{fig2}}
\end{figure}

\begin{figure*}[t]
\epsscale{2.0}
\plotone{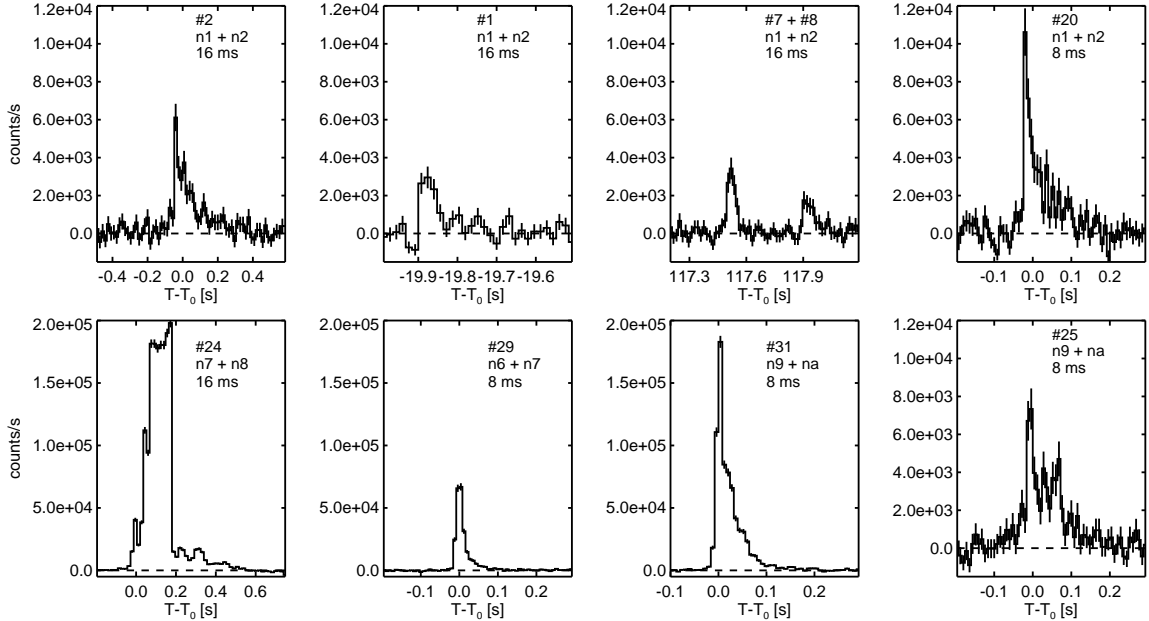}
\caption{Background subtracted light curves of \sgr~ bursts observed during the 2008 October (top row) and the 2009 March -- April activity periods (bottom row). Top row: the left three panels show the triggered burst of Figure 2 (at $T_0 =  $ 09:03:06 UT, 244717387 MET) and three untriggered events, namely burst \#1 at $T_0  - 19.9$ s and bursts \#7 \& \#8 at $T_0 +  117.5$~s. The rightmost panel shows the brightest burst (\#20) observed during the 2008 October period. Bottom row:
the left three panels show the brightest bursts ( \#24, \#29 and \#31) observed during the 2009 March -- April period. The rightmost panel shows burst \#25 with comparable intensity to burst \#20 shown in the panel above. The light curves were obtained by adding counts of the two detectors (depicted in the panels in hexadecimal notation, n0 $...$ nb), which showed the highest peak count rates. \label{fig3}}
\end{figure*}

\section{Data analysis}\label{sec:analysis}

The \fe/GBM instrument comprises two different types of scintillation detectors.  Burst locations and the low energy part of spectra (8 keV to 1 MeV) are provided by an array of 12 Thallium-activated sodium iodide (NaI) scintillator detectors (numbered from NaI~0 to NaI~11). The two Bismuth Germanate (BGO) detectors are extending the spectral coverage to $\sim 40$~MeV \citep[for a detailed overview of the instrument and its capabilities see][]{mee09}. Here we use only the spectral data below 200~keV, since none of the examined bursts showed emission at higher energies.

GBM provides several trigger data types: the continuous high spectral resolution data CSPEC, with 128 energy channels, logarithmically spaced over the energy range of the NaI detectors and  1.024~s time resolution; the continuous high time resolution CTIME data, binned at 64~ms, with only 8  channel spectral resolution; and the Time-Tagged Event (TTE) data, which we used  for the analysis of the relatively short SGR bursts. The TTE data consist of a time-tagged photon event list with a temporal resolution as low as 2~$\mu$s and the same spectral resolution as the CPSEC data. The TRIGDAT data, the first burst information, quickly downlinked after a burst trigger, are comprising information on background rates, the burst intensity, hardness ratio, localization, and classification, which is needed for the on-ground localization.

\subsection{Data Selection Criteria}

\fe~ is operated in sky-scanning survey mode, with the viewing direction rocking to North and South by $35^{\circ}$  from the zenith on alternate orbits, at a slew rate up to several degrees per minute. This operation mode causes a continuous change of the angle between the source and the detector normal. We have analyzed here bursts viewed by detectors with source aspect angles less than $40^{\circ}$. This selection is justified by the softer SGR spectra and the detector effective area, which is decreasing more rapidly with source angle for softer photons (\citealt{bis09}). None of the selected detectors was subject to blockage  from other parts of the \fe~ spacecraft.

Figure~\ref{fig2} shows a 200\,s section of the TTE data light curve during the first  trigger bn081003.377 in 64\,ms bins. As indicated on the figure, we find eight untriggered bursts during this period, in addition to the one that triggered the instrument at $\sim 0$~s. Table~\ref{tbl-1} lists the start times of all individual bursts related to the trigger time, together with the detector numbers and data types used in our analysis. In this first trigger, detectors NaI 1 \& 2  had the smallest source aspect angle during the $\sim330$\,s TTE data period. At times starting after the TTE data type ends until the end of the trigger mode at  $\sim~600$~s, only continuous CSPEC \& CTIME data are recorded, and detector NaI 5  also fulfills the $<40^{\circ}$ selection criterion.

Figure~\ref{fig3} exhibits individual bursts from the two active periods. The first panel of the top row shows trigger bn081003.377, which triggered GBM at $T_0 =$~9:03:06~UTC on 2008 October 3 (Burst \# 2 of Table~\ref{tbl-1}; see also Figure~\ref{fig2}). The next two panels show untriggered bursts: the first event at $T_0 \sim - 19.9$~s, which was too weak to trigger GBM, and bursts \# 7 \& \#8, observed at  $T_0 + 117.5$~s. The light curve of the brightest event during this period, Burst \# 20, which triggered GBM one day later, is shown in the forth panel. The shape, intensity and duration of all these events are typical for all other bursts observed during the 2008 October activity period.

During the 2009 March -- April period only 15 bursts (also listed in Table~\ref{tbl-1}) were observed, including two very bright bursts with single detector peak count rates
$> 1 \times 10^5$~cnt/s. At these high count rates the performance of GBM is affected by dead time and pulse pile-up as described in detail in \cite{mee09}.
The light curve of trigger bn090322.944 (Burst \# 24 of Table~\ref{tbl-1}, shown in the first, bottom row panel of Figure~\ref{fig3}), saturated GBM for  $\sim 100$~ms.
The flat plateau from $T_0 + \sim 80$~ms  to  $T_0 + \sim 185$~ms  is caused by the clipping of TTE events at the maximum rate of the High Speed Science Data Bus (HSSDB). Burst \# 29 (trigger bn090330.237), shown in the second panel, reached a single detector peak count rate of about $> 3.5 \times 10^4 $~cnt/s,  resulting in dead time of $10 \%$. The third panel shows the second brightest burst (Burst \# 31 of Table~\ref{tbl-1}) with an observed single detector peak count rate of $\sim 1 \times 10^5$~cnt/s and a dead time of $ 50\%$.  The non saturated bursts of the 2009 March -- April period are on average brighter than the bursts observed in 2008 October. The right most panel of the bottom row shows Burst \# 25, which is comparable in brightness to the brightest burst observed during 2008 October.

Table~\ref{tbl-1} lists detailed trigger information for all events as well as the time periods (Start- \& Stop-columns), used for our spectral analysis. The high temporal resolution of the TTE (2 $\mu$s) easily accommodates the analysis of the bursts. The 0.064~s continuous CTIME data cannot be used for our spectral analysis, due to their coarser spectral resolution. CSPEC data were used only for burst \# 15 (Table~\ref{tbl-1}), which was located inside the one CSPEC 1.024 s integration time with a signal to background ratio high enough to allow spectral analysis (this burst was excluded from temporal analysis).

\subsection{Soft Burst Location Systematics}\label{sec:burstlocations}

We determined for all 21 triggered bursts in Table~\ref{tbl-1} an approximate on-ground location using the TRIGDAT-data, to check the consistency of  the GBM derived locations with the accurately known source position. These locations were computed using the Daughter Of Locburst (DOL) location-finding code, which is comparing observed rates to a table of calculated relative rates in the 12 NaI detectors for each of 41168 directions (at 1 degree resolution) in spacecraft coordinates by utilizing a chi-squared fit. Compared to the on-board flight software locations, the DOL is including atmospheric and spacecraft scattering more accurately, using a finer angular grid and is more properly accounting for differences in burst spectra. In addition DOL also corrects for the recording dead time. A dedicated table was prepared especially for soft events with rates integrated over the $5 - 50$~keV energy range.

\begin{figure*}[t]
\plotone{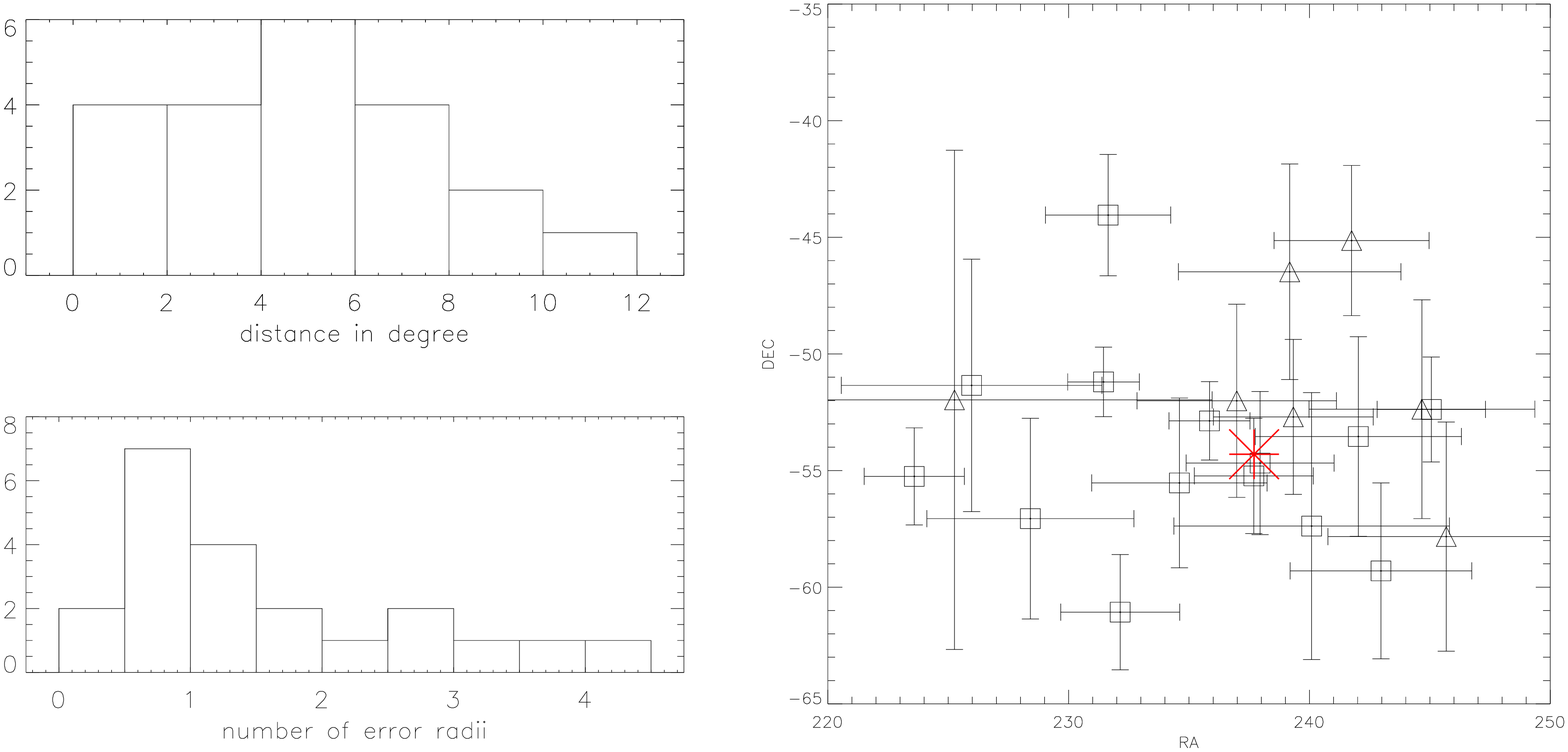}
\caption{Right panel: On-ground calculated locations and their uncertainties (radius of 1$\sigma$ containment). The locations from the two activity periods are marked with different symbols (2008 October: open triangles, 2009 March -- April: open squares). The source position of \sgr~is shown as a cross.
Left panels: (top) Location accuracy, displayed as absolute distance in degrees from the source position, and (bottom) as distance in number of error radii.
\label{fig4}}
\end{figure*}

\begin{figure*}[t]
\plotone{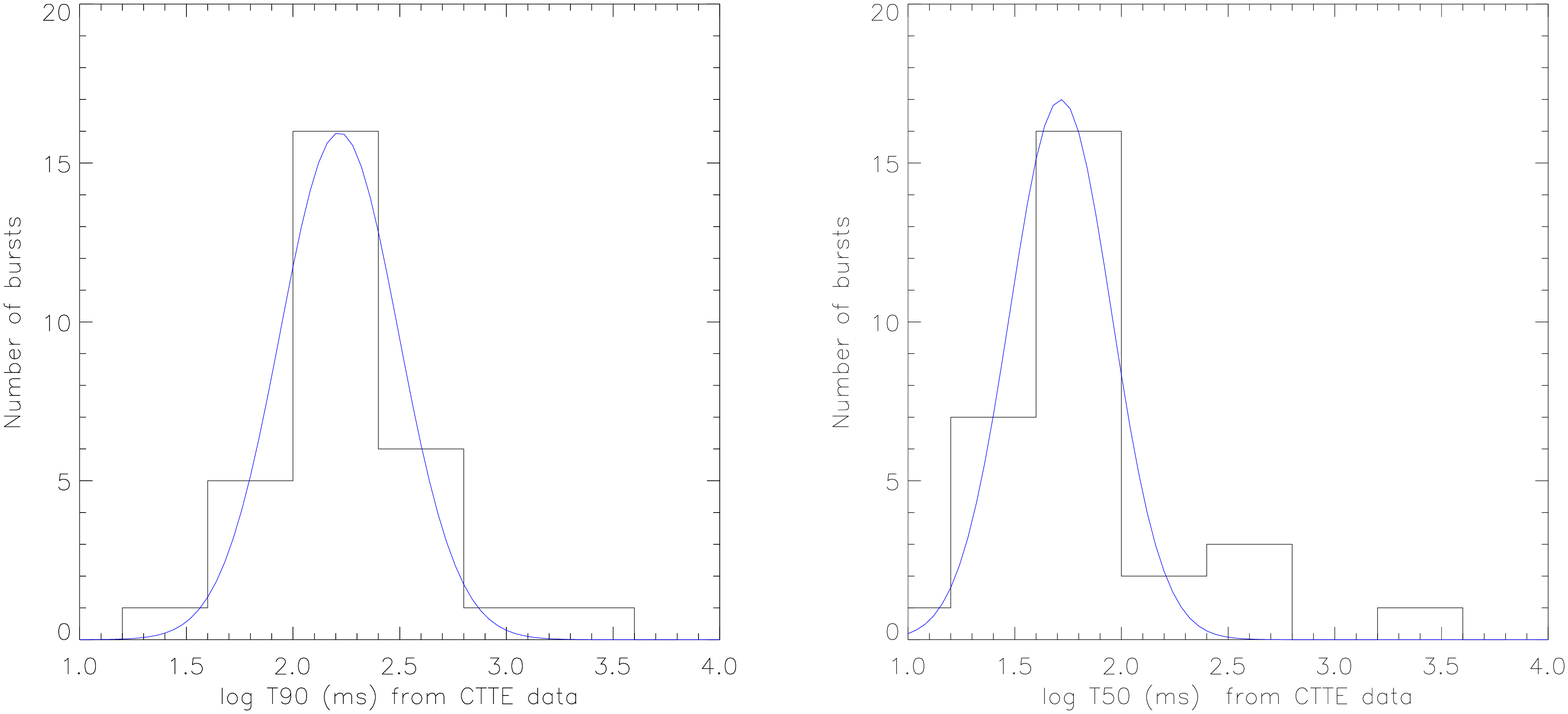}
\caption{ Distribution of the $T_{90}$ (left panel) and $T_{50}$ (right panel) CTTE durations, fit with log-normal functions. The means / standard deviations of the $T_{90}$ and $T_{50}$ distributions are $2.22 \pm 0.02$  ( $= 165 \pm 8$~ms) /  $ 0.28 \pm 0.02$ and $1.72 \pm 0.04$  ($ = 52 \pm 5$~ms) / $0.24 \pm 0.03$ , respectively. The normal mean values of both distributions are 278~ms and 138~ms, respectively.
\label{fig5}}
\end{figure*}

The right panel of Figure~\ref{fig4} shows the calculated locations and their statistic uncertainties (radius of 1$\sigma$ containment) for both activity periods. The two left panels are showing the location accuracy distributions, as absolute distance in degrees from the source position (top), and as distance in number of 1$\sigma$ radii (bottom). Roughly 86\% of the locations are consistent with the source position within 3$\sigma$ error radii. A similar percentage are located $<8$\degrees~ from the actual source position in a flat distribution. It should be noted that the three bursts (\#31, \#45 \& \#36)  with the largest deviation of error radii ($> 3$), are the ones with the highest 16~ms peak fluxes, listed in Table~\ref{tbl-2}, with the exception of trigger bn090322.944, the saturated burst \#24 (see Figure~\ref{fig3} bottom row, leftmost panel). In this case the location was derived for one selected $\sim 50$~ms bin in the rising part of the burst lightcurve, from -0.048~s to 0.016~s, thus avoiding the saturated part. From these dependencies we conclude that the 'soft' location accuracy is affected by high event rates, probably caused by hidden systematic errors, e.g.~the standard dead-time correction may not be valid for soft \& bright events.

The accuracy of the GBM localizations for (the hard) Gamma-Ray Bursts (GRBs) is reported by Connaughton et al. (2012, in preparation). They find that the systematic location errors for GRBs are best described by two contributions, one with 2.8\degrees~ error with 70\% weight and one with 8.4\degrees~ error with 30\% weight.

\subsection{Temporal Analysis}\label{sec:TempAnalysis}

\subsubsection{$T_{90}$ and  $T_{50}$ durations}

The \sgr~burst temporal analysis was performed in a similar manner as the one described by \cite{Lin11}. The durations, expressed as $T_{90}$ and  $T_{50}$ \footnote{$T_{90}$ ($T_{50}$) is the duration during which the background- subtracted cumulative counts increase from 5\% (25\%) to 95\% (75\%) of the total counts \citep{kou93}.}, were determined both in photon and in count space. The durations in count space were calculated by applying an algorithm adapted for the analysis of GBM TTE data, which was originally developed by \cite{kou93} and later modified by \cite{gog01} and  \cite{gav04}. Thanks to a new on-ground build data type of the GBM data, the so-called CTTE data type, it was possible to determine the $T_{90}$, $T_{50}$-durations in photon space using the GBM {\it RMFIT (4.0rc01)} software\footnotemark{}\footnotetext{R.S.~Mallozzi, R.D.~Preece, \& M.S.~Briggs, "RMFIT, A Lightcurve and Spectral Analysis Tool," \copyright 2008 Robert~D.~Preece, University of Alabama in Huntsville, 2008}, similar to what was done in the case of GBM GRBs \citep{paciesas2012}. The CTTE data type is  derived from TTE data by rebinning the 128 TTE energy bins into 8 energy bins, with the same boundaries as the CTIME data. The CTTE data type allows to generate finer time bins,  compared to 64 ms bin width of the CTIME data during burst mode, which are necessary especially for temporal analysis of short bursts from SGRs. In our case the individual burst data were rebinned to 4~ms, 8~ms or 16~ms bins depending on the intensity of the burst. The burst durations $T_{90}$ ($T_{50}$) were computed in the $8 -200$ keV energy range.

Since the bursts  presented here are relatively weak, we applied the count rate algorithm twice to each burst in order to account for systematics. In some cases the deviations were large, most probably depending on small differences in the background selection. Durations calculated in photon space have smaller systematic uncertainties. Due to this fact, only the results  obtained in photon space are presented here.

Figure~\ref{fig5} shows  the distribution of $T_{90}$  and $T_{50}$ durations for 30 bursts, out of the 37 bursts listed in Table~\ref{tbl-1}. The mean value of $\sim 170$~ms of the log-normal fit to the $T_{90}$ distribution is comparable to the bulk of magnetar bursts and entirely consistent with the average duration of $174\pm10$~ms of the bursts during the most active episode from the source (2009 January) reported by van der Horst et~al. (2012). No general trends or differences in the burst durations of the two active periods was observed.

\subsubsection{Timing Analysis}\label{sec:TimingAnalysis}

In 2008 October \sgr~emitted 19 bursts in the course of $\sim10$~hours. Out of these bursts, 15 events were emitted within only $\sim10$~min.  Similarly to \citet{kan10}, we looked for periodic modulations in the source persistent emission using the TTE data stream of NaI~1, which was the detector with the smallest source angle towards \sgr. For the analysis we corrected the TTE data to the solar system barycenter, binned the data with a time resolution of 16~ms ($8-100$\,keV), and selected the time interval $[T_0-25; T_0+160]$~s, which roughly corresponds to the interval from burst \#1 to \#9 (see Table~\ref{tbl-1}). We performed all timing analyses twice, with the SGR bursts and without them (we carefully removed the burst intervals from the light curve).

We used several different timing techniques, including a Lomb-Scargle periodogram \citep{sca82, horne86, press89}, \textit{Period04}\footnote{http://www.univie.ac.at/tops/Period04/} \citep{lenz05}, a code specifically designed to extract multiple periodic signals in astronomical time series through simultaneous least squares fitting. We also used a multi-harmonic ``analysis of variance'' periodogram (mhAoV) of \citet{schw96}, and the method introduced by \citet{vaughan05}. After correcting for the number of trials, no significant frequency was found in either of the data sets.

An intriguing finding was in the burst arrival times, namely the time separation between bursts. Starting from the first event, 8 out of 16 bursts occur with a time separation in multiples of $10\pm0.5$~s. It is established observationally that SGR bursts occur randomly distributed in rotational phase (Mereghetti 2008). We performed a simple Monte Carlo simulation to test the significance of this result. Our simulations showed that the observed burst interval corresponds to a single-trial probability of  $P=6.1\times10^{-5}$, i.e., a significance of $\sim 4 \sigma$. It is hard to accurately quantify a number of trials a posteriori. Considering the fact that there are various SGR/AXP sources, with several multiple prolific bursting episodes, the probability will be in general smaller. Since the January emission was heavily concentrated in a 24 hour period with a very large amount of bursts, we have not performed a detailed search for the 10~s period during this epoch. Only future observations will reveal whether this periodicity is indeed real and related specifically to \sgr, or simply an observational artifact.

\begin{figure*}[t]
\plottwo{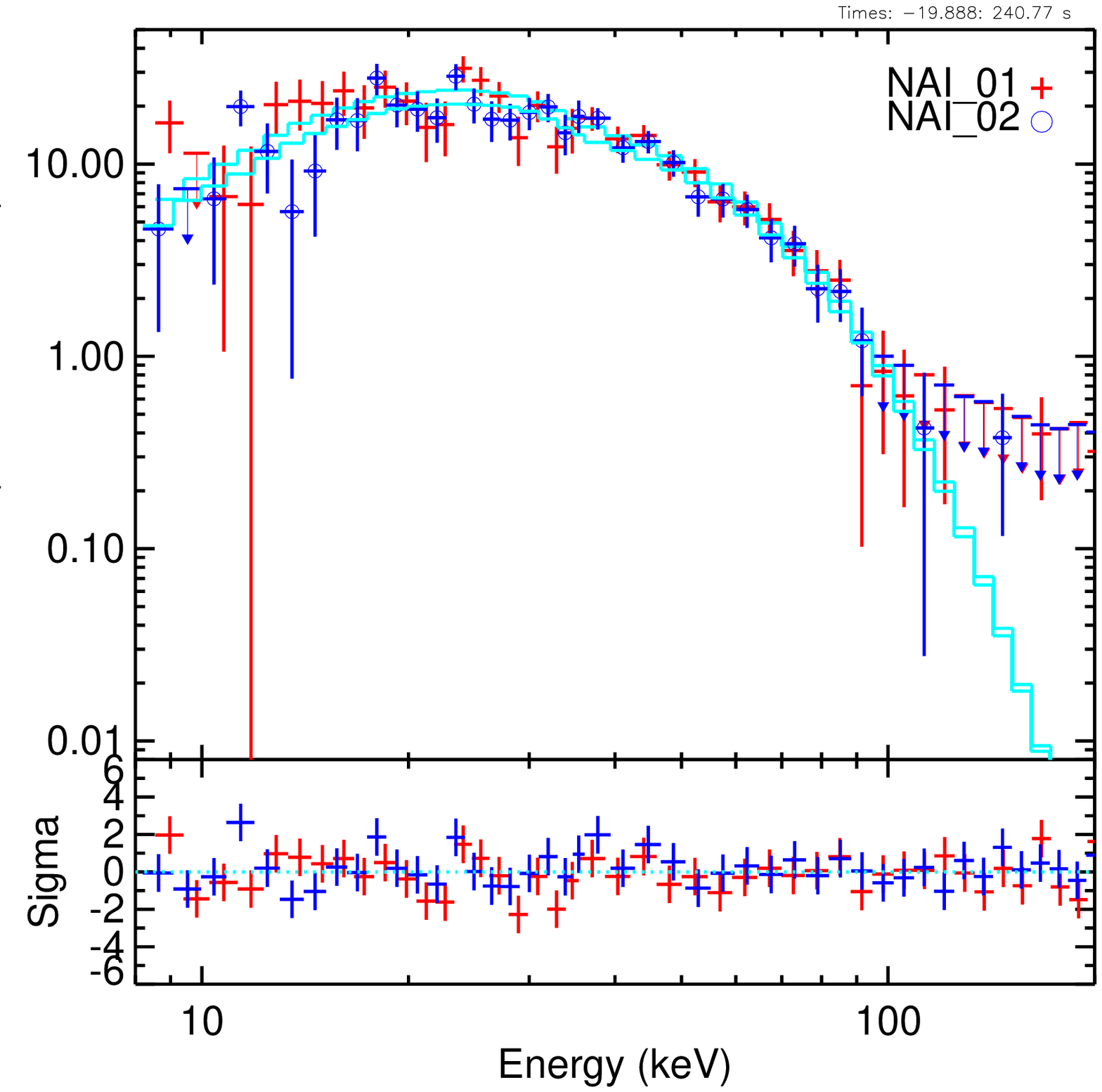}{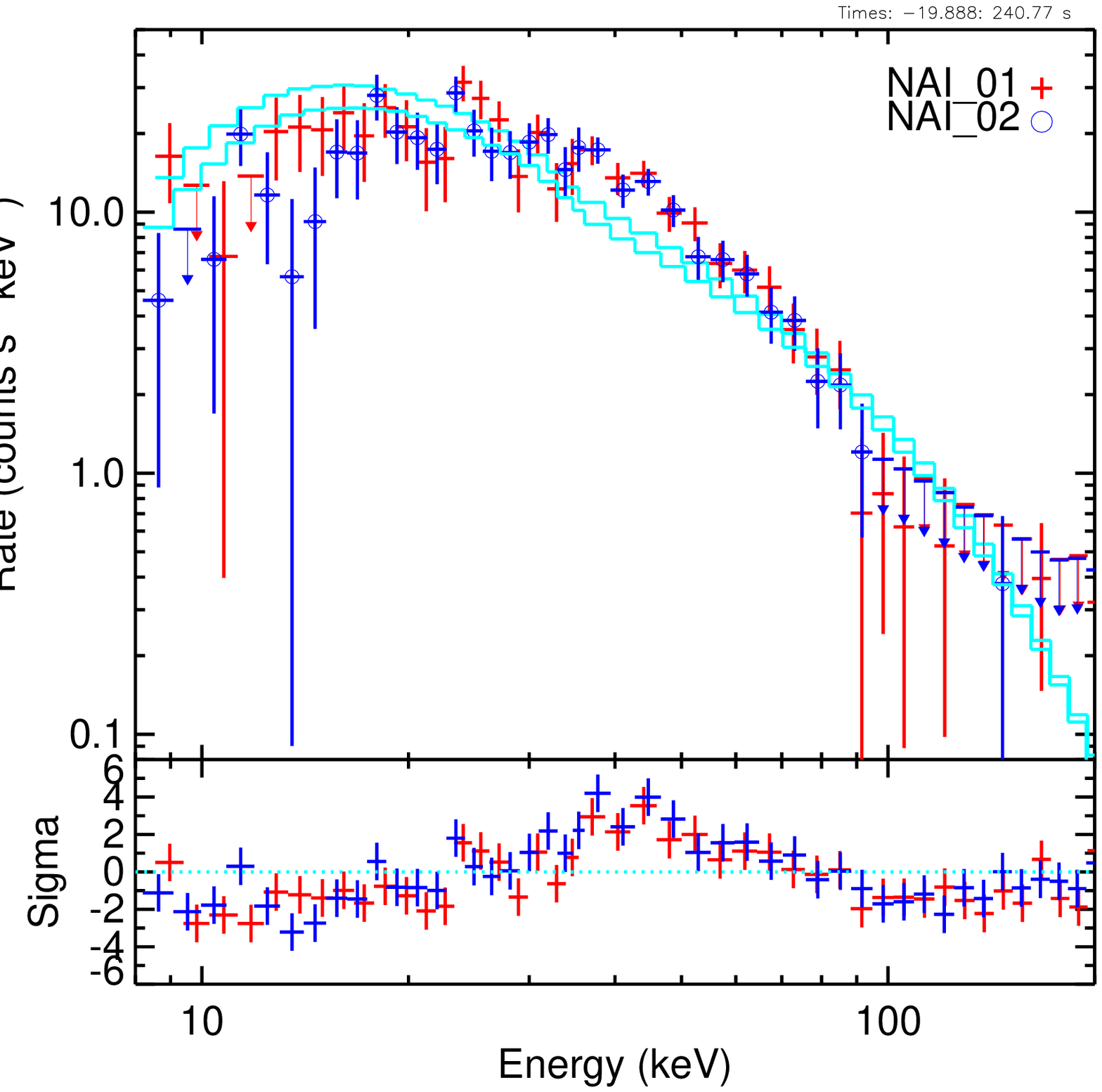}
\caption{Stacked spectra of bn0810030.377 Left: BB Fit, Right: OTTB Fit \label{fig6}}
\end{figure*}

\begin{figure*}[t]
\plottwo{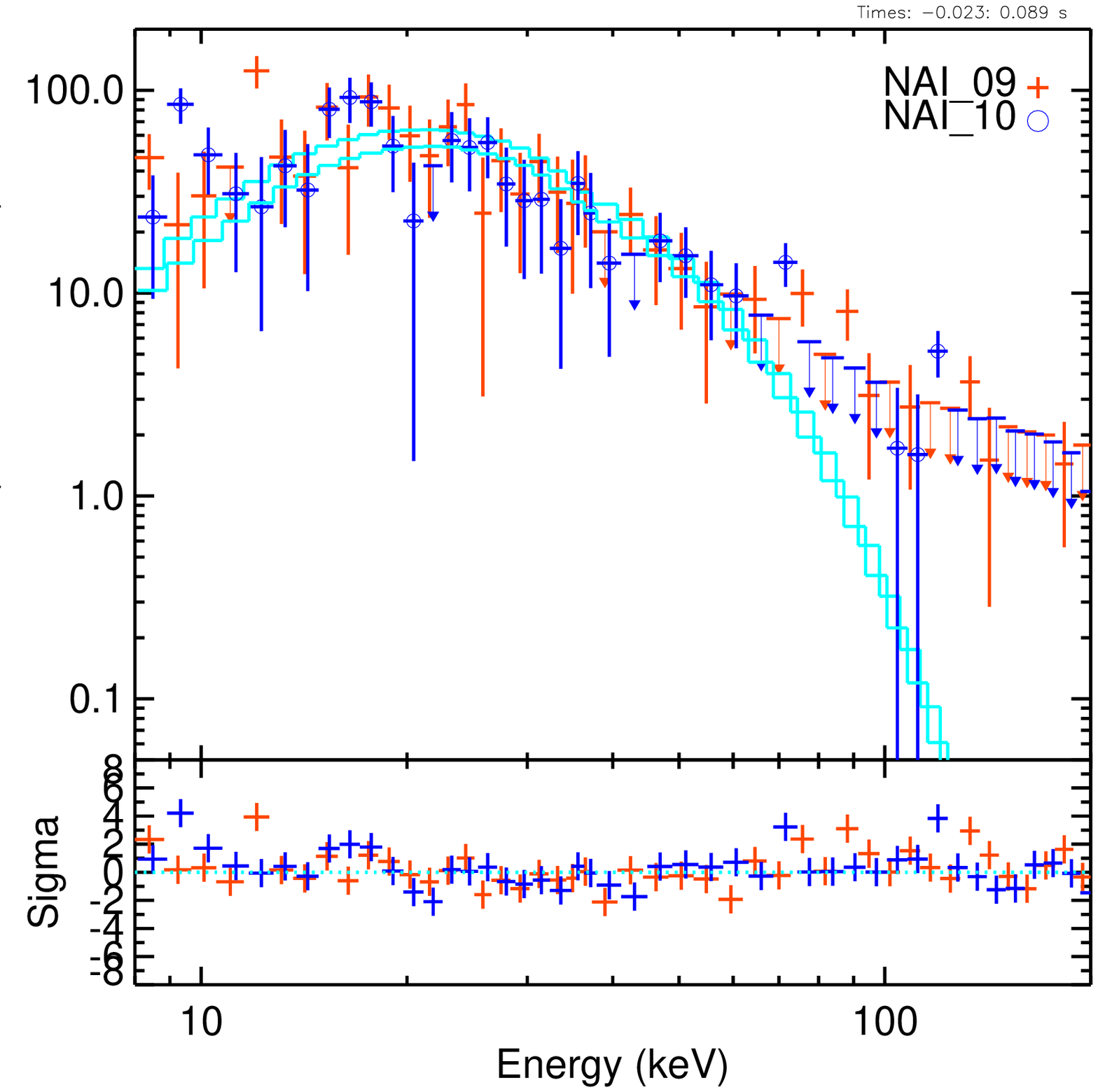}{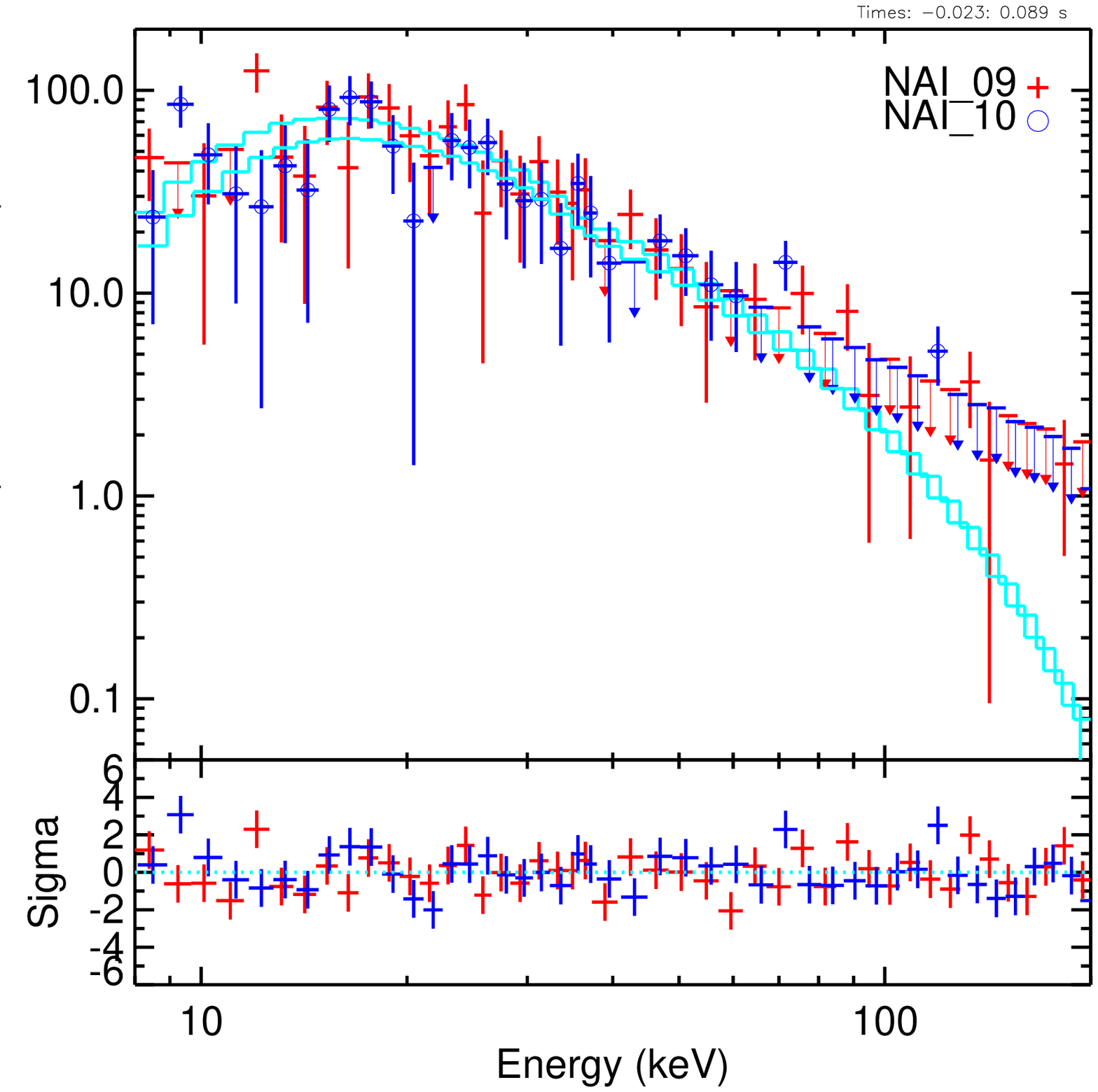}
\caption{Spectra of bn090326.625. Left: BB Fit, Right: OTTB Fit \label{fig7}}
\end{figure*}

\subsection{Spectral Analysis}\label{sec:SpecAnalysis}

We fit all bursts in Table~\ref{tbl-1} with the following models, which are known to best approximate SGR spectra: a simple power law function (PL), a power law function with an exponential high-energy cutoff  (Compt), a blackbody (BB) and an optically thin thermal bremsstrahlung (OTTB) spectrum and a combined BB$+$BB model. We used the {\it RMFIT (4.0rc01)} spectral analysis software developed for the GBM data analysis. Table~\ref{tbl-2} lists the parameters of all single models (with the exception of the PL model) for both activity periods.

Combining all our fit results we find that the 2008 October burst spectra are best fit with a single BB model. This is corroborated by the fact, that the brightest burst observed during that period (burst \# 20 of Table~\ref{tbl-1}) as well as a stacked spectrum derived from all bursts observed during the TTE data period of the bn081003.377 trigger (Burst \#1 to \#10 of Table~\ref{tbl-1}) are  best fit by the BB model as compared to fits with the OTTB and PL models. The BB and OTTB fits to the stacked spectrum are shown in Figure~\ref{fig6}. We note that the residuals of the OTTB-fit show a systematic wiggle, causing a $\Delta$C-stat between the two models of $\sim 165$. The same deviations were observed for burst \#20, this time with a $\Delta$C-stat of $\sim 40$. A fit with the Compt-model yields in both cases a C-stat comparable to the value obtained for the BB fit, but the Compt model has one degree-of-freedom (DOF) less. For most of the bursts during the 2008 October period the $\Delta $C-stat for the BB model was $> 6$ with the exception of some weak bursts, which were also well fit by an OTTB or PL model.
A fit with a normal function to the distribution of single BB temperature $kT$ yields a mean value
of $12.4 \pm 0.2$~keV (width, $0.9 \pm 0.1$~keV).
The preference for a single BB model as best fit model was also reported by \cite{isr10} for bursts observed during 2008 October with \swift-BAT in the $15 - 100$~keV energy range. These bursts are all well described by a single BB function with temperatures $\sim 10$ keV.

We also tried to fit the 2008 October bursts, even though they are faint, with the combined two BB functions, since \cite{Lin12} reported that broadband (0.5 - 200 keV)\swift-XRT/GBM spectral fits are showing, on average, that the burst spectra are better described with the two BB functions than with the Comptonized model. As was expected, these fits were not able to constrain simultaneously the parameters of both BB components, even when fixing the ratio of the fluence in the two BB components, so that the hot component has twice the flux of the cold component, which is the mean ratio with relatively modest scatter that is found in \cite{Lin12}.

In contrast, the bursts from the 2009 March -- April period are best fit in almost all cases by an OTTB-model. In order to investigate whether this difference is connected with the burst brightness difference between the two periods, we selected a burst with comparable brightness from each period, as shown in the rightmost panels of Figure~\ref{fig3}, the brightest of the 2008 October activity period, already presented above (burst \# 20 of Table~\ref{tbl-1}), as well as  a burst among the fainter ones of the 2009 period (burst \# 25 of Table~\ref{tbl-1}).
The results of the  BB, OTTB and Comp model fits for burst \#25 are listed in Table~\ref{tbl-2} , showing indeed that the OTTB is the preferred model, with an improvement in C-stat of $> 30$. The BB and OTTB spectral fits are shown in Figure~\ref{fig7}. Finally, the spectral analysis of the bursts with comparable flux values ($\sim 50$~ph/cm$^2$/s) and fluences ($\sim 13\times 10^{-8}$~erg/cm$^2$), yielded significant differences of $\Delta$C-stat $> 30$ for the best-fit models, leading us to the conclusion that this difference in spectral shape is intrinsic, and probably caused by a change in the burst emission process during the two periods.

We would like to point out that the Compt and BB+BB models fit the burst spectra of this activity period equally well, with the exception of some events where fit parameters remained unconstrained. Using the successful BB+BB fit results we obtain a mean value for the cool BB temperatures of $5.1  \pm 0.1$ keV (width, $0.6  \pm 0.1$ keV), and for the hot BB temperatures of $14.7 \pm 0.2$ keV (width $0.9 \pm 0.1$ keV). Only in the case of the second brightest burst \#31, the best fit model was the Compt-model, with a $\Delta$C-stat of $\sim 40$, compared to an OTTB fit, the BB+BB model in this case did fit equally well. This result is not conclusive, since the spectral slope of bright bursts could be affected by pulse pile-up effects as already pointed out in \cite{AvdH12}.
The spectral analysis of the brightest burst \#24 was performed by excluding the saturated parts; nevertheless, none of the models, including the combined ones, provided a reasonable fit. Most probably the whole emission period was affected by pulse pile-up and saturation effects.

To identify the model that best describes the \sgr~ burst data, we performed simulations similar to \cite{AvdH12} with {\it RMFIT (4.0rc01)}. We selected the two bursts with similar brightness from the October and March -- April  periods, burst \#20 and \#25, respectively. For each detector and each event a set of 10000 synthetic spectra were created. The background counts of these spectra are estimated from the real data, whereas the source counts are computed from the function which was used to fit the real data, folded with the detector response matrix (DRM). Poissonian noise was added to the sum of the source and background counts. During the fit process a synthetic background spectrum, with added Poisson fluctuations to each energy channel, was subtracted from the synthetic burst spectrum.

For burst \#20 a set of 10000 synthetic spectra were created with the OTTB function and its best parameters from the fit to the real data were used as null hypothesis. The 10000 spectra were then fit with both the OTTB and a single BB function. The distribution of the difference in C-stat ($\Delta$C-stat) was then compared to the $\Delta$C-stat obtained from the real burst data, which is $\Delta$C-stat$_{OTTB-BB}=42$. There is not a single synthetic burst which exceeds this difference. In fact, the highest $\Delta$C-stat$_{max}=6.3$. Subsequently, we conclude that statistical fluctuations cannot account for the difference in the statistic between OTTB and BB. Therefore, the null hypothesis is rejected ($P < 10^{-4}$) and we conclude that the BB function is the preferred model for this emission epoch.

The same line of reasoning was applied to the 2009 March event. However, contrary to the burst above, this time the BB was taken as the null hypothesis, i.e., 10000 synthetic spectra were created using the BB model and its best fit parameters from the real data as input model. For burst \#25 we observe $\Delta$C-stat$_{BB-OTTB}=36$ in the real data. However, the maximum value of the simulated C-stat distribution is 10. Similarly to what is observed above, we conclude that statistical fluctuations cannot account for the difference in C-stat and therefore we find the OTTB function to be the preferred model for this event.

\section{Discussion}\label{sec:discussion}

Our main new finding in this work is that the observed GBM spectrum of the bursts in the energy range $8-200\;$keV has changed from being BB-like in the 2008 October active period to a broader, steeper OTTB-like form with less curvature during the 2009 March  activity episode (see Table~\ref{tbl-2}). In particular, the photon index below the peak, $\lambda$ that is defined through $dN/dE \propto E^\lambda$ has changed from $\lambda \sim 1$ during 2008 October to $\lambda\sim -1$ during 2009 March .  In this context, it is interesting to note that during the most active bursting period of \sgr, in 2009 January , the GBM burst spectra were typically not well fit by a single BB spectrum, but were instead equally well fit by an OTTB, Comptonized or
BB+BB spectrum \citep{AvdH12}. When including also
Swift/XRT data in the spectral fit of those 2009 January bursts for which these were available, however, a BB+BB spectrum is usually preferred \citep{Lin12} in which the cool BB component has a factor of $\sim 2$ or so smaller fluence than the hot BB component (interestingly enough, when fit to a Comptonized spectrum, the implied values of $\lambda$ typically ranged between $\sim -1$ and $\sim 0$). The 2008 October are bursts best fit by a BB component of temperature $\sim 11-14\;$keV and effective area $\sim 0.2-2\;$km$^2$, which are similar to the hot BB component found for the 2009 January bursts. It is possible that a cool BB component could potentially still be present in the 2008 October bursts. If we assume a cool BB to hot BB flux ratio similar to that of the 2009 January bursts, then in order to avoid clear detection in the GBM spectra presented here the cool BB temperature typically needs to be below a few keV, which would correspond to an effective area of $\gtrsim 10^3\;$km$^2$.  In this scenario both the hot and cold BB components in the 2008 October bursts would be near the cool end of the corresponding components from the BB+BB spectral fits for the 2009 January bursts.

The differences between the spectroscopy of the bursts studied here and for the bursts in 2009 January clearly indicates evolutionary patterns on the timescale of a few months or so. This behaviour might arise, for example, from a change in the details of the energy release or containment between different episodes.  The rather small effective areas we obtain for the BB spectrum ($\sim 0.2-2\;$km$^2$) imply a small emission region, which is therefore probably close to the NS surface. Such a proximity to the surface would therefore tend to result in a relatively high effective opacity, due to large plasma densities.  The opacity is mainly controlled by the scattering cross section of the ordinary or O-mode photons being near the Thomson value.  O-mode photons are those where the photon electric field vector lies in the plane defined by their momenta {\bf k} and the local magnetic field. Photons in the extra-ordinary polarization mode (or E-mode, where the photon electric field vector is normal to the local {\bf k}--{\bf B} plane), experience a dramatically reduced scattering cross section that is suppressed because the photon energy is typically far below the cyclotron energy (e.g., \citealt{her79}).  These E-mode photons contribute little to the overall opacity, which being high, results in a local quasi-thermodynamic equilibrium (i.e., LTE) that should drive the spectrum towards a BB or BB+BB form.

Radiative transfer effects in a strong magnetic field ($B\gg B_{\rm QED}$) can cause a deviation from a pure BB spectrum  (e.g., see  \citealt{ulmer1994,lyubarski2002}) due to the increase in E-mode opacity with photon energy, that enables lower energy (E-mode) photons to escape from a larger depth within the emission region, where the temperature (that is established by the O-mode photons) is higher, thus resulting in a softer than thermal spectral slope below the peak, where $dN/dE\propto E^\lambda$ with $\lambda \sim 0$.  Somewhat harder low-energy spectral slopes might still be possible, e.g., due to resonant ion cyclotron absorption, which should be sensitive to the local value of the magnetic field. In order for such an absorption to reach sufficiently high photon energies, a rather high local value of the magnetic field is required in the emission region ($\gtrsim 10^{15}\;$G), which for \sgr~is in excess of the surface dipole magnetic field strength (of $\sim 2.2\times 10^{14}\;$G) inferred from its $P\dot{P}$.  Thus, this naturally suggests higher multipoles (e.g.,\citealt{tho02}). A strong local magnetic field from quadrupole and higher multipole configurations could effectively confine the hot emitting plasma to relatively small closed magnetic flux tubes near the stellar surface.  This is consistent with the small effective areas inferred from our spectroscopic analysis. \footnote{Some hint for the presence of higher multipoles might be found in the irregular shape of the pulse profiles of \sgr, which change as a function of energy and time (e.g., \citealt{kan10,Lin12}), but this is by no means conclusive.} Thus, the spectral slope below the peak energy might be related to the local field strength and topology near the magnetic dissipation region that gives rise to the bursting episode. This might potentially be the factor that is common to different bursts within the same active period, but varies between different activity episodes, i.e. field topology evolves significantly on timescales of a month or so.
It is also possible that the trigger from the crustal regions may relocate to different colatitudes during this evolution, thereby precipitating a sampling of disparate
field topologies within the magnetospheric dissipation zones involved.

The broader OTTB-like spectra of the 2009 March bursts might reflect a Comptonized spectrum from an emitting region with a modest optical depth.  As discussed, e.g, in  \cite{Lin11,Lin12} \citep{1979rpa..book.....R}, the simplest form of such a model yields $\lambda = 1/2-\sqrt{9/4+4/y_B}$ where $y_B= 4kT_e/(m_ec^2)\max[\tau_B,\tau_B^2]$
is the magnetic Compton y-parameter, $\max[\tau_B,\tau_B^2]$ is the mean number of scatterings per photon by the hot electrons, and $\tau_B$ is the effective optical depth for scattering that in our case is significantly modified by the strong magnetic field (and is thus dubbed the magnetic optical depth).  For \sgr, the inferred peak energies for the 2009 March bursts, typically $E_{\rm peak}\sim 30-45\;$keV, suggest $4kT_e/(m_ec^2) \sim 0.23-0.35$, which would imply $\lambda \sim -0.8$ and $\sim -0.95$ for $\tau_B \sim 5$ and $\sim 10$, respectively.\footnote{Note that relativistic corrections for such large temperatures, such as Klein-Nishina reductions, only influence these inferred indices to a modest extent; see \cite{Lin11} and references therein.} Thus such modest values of $\tau_B$ would result in $y_B\gg 1$ so that $\lambda$ approaches the value of $-1$, approximately coinciding with the lower energies of the inferred OTTB-like spectrum of the 2009 March bursts. Much larger optical depths would result in saturated Comptonization or true thermalization, and a spectrum closer to a BB, though still generally different from a BB as mentioned above.
Moreover, the 2009 March bursts are equally well fit by a BB+BB spectrum, so that in principal it is possible that the underlying spectrum during all the three bursting periods discussed above (2008 October , 2009 January and 2009 March ) are in fact BB+BB or multi-blackbody, which was discussed in detail in \cite{AvdH12}.

Different bursts within the same activity period may exhibit a relationship between the spectrum and the timing of the bursts.  A given active period might be triggered by the yielding of the crust to magnetic stresses at a particular location on the NS. The magnetic field structure in that region could affect the details of the energy release and confinement of the hot plasma that is produced, and thereby influence the resulting spectrum of the bursts. If the emitting region is small and near the surface then it might be obscured during certain rotational phases (since the burst duration is smaller than the rotational period), hence resulting in a non-uniform distribution of bursts with the rotational phase, as was found by \cite{Lin12}.
If the bursts indeed span a reasonable range of rotational phases, then it is likely that they sample substantially different viewing angles with respect to the well-localized emission region. This would then suggest that the viewing angle is not the dominant factor in determining the overall spectral shape.

\acknowledgments

This publication is part of the GBM/Magnetar Key Project (NASA grant NNH07ZDA001 -GLAST, PI: C. Kouveliotou). Support for the German contribution to GBM was provided by the Bundesministerium f\"ur Bildung und Forschung (BMBF) via the Deutsches Zentrum f\"ur Luft und Raumfahrt (DLR) under contract number 50 QV 0301. A.v.K. was supported by the Bundesministeriums f\"ur Wirtschaft und Technologie (BMWi) through DLR grant 50 OG 1101. C.K. and A.J.v.d.H. were partially  supported by NASA grant NNH07ZDA001-GLAST. M.G.B. acknowledges support from NASA through grant NNX10AC59A.
E.G. and Y.K. acknowledge the support from the Scientific and Technological Research Council of Turkey (T\"UB\.ITAK) through grant 109T755. L.L. is supported through the Post-Doctoral Research Fellowship of the Turkish Academy of Sciences (T\"UBA). D.H. and A.L.W. acknowledge support from an NWO Vidi grant (PI: A.L.~Watts)

\clearpage
\begin{deluxetable}{clrrrrrcc}
\tabletypesize{\tiny}
\tablecaption{\sgr~ trigger times \& durations \label{tbl-1}}
\tablewidth{0pt}
\tablehead{
\colhead{Burst} &
\colhead{Trigger\tablenotemark{a}} &
\colhead{Trigger time} &
\colhead{Start\tablenotemark{b}} &
\colhead{Stop\tablenotemark{b}} &
\multicolumn{2}{c}{Duration} &
\colhead{Detectors} &
\colhead{Data} \\
\# & \colhead{\#} &  \colhead{UTC} & & &\colhead{T90} & \colhead{T50} &\colhead{\#} &\colhead{Type}   \\
 & & & \colhead{(s)} &	
\colhead{(s)} &	
\colhead{(ms)} &		
\colhead{(ms)} & &  }
\startdata
 1	& bn081003377U &  9:03:06 &	-19.888 &	-19.840 &	$160 \pm 29$ & $72 \pm 11$ &	1, 2 &	TTE/CTTE  \\	
 2	& bn081003377 &  9:03:06 &	-0.048 &	0.128	&   $276 \pm 53$ & $120\pm 25$ &1, 2 &	TTE/CTTE  \\
 3	& bn081003377U &  9:03:06 &	10.112 &	10.496 &	$672 \pm 72$ & $320 \pm 72$	& 1, 2 &	TTE/CTTE  \\	
 4	& bn081003377U &  9:03:06 &	27.168 &	27.264 &	$68\pm 25$ & $28 \pm 11$	&1, 2 &	TTE/CTTE  \\	
 5	& bn081003377U &  9:03:06 &	89.568 &	89.664 &	$128 \pm 23 $ & $56 \pm 23$	&1, 2 &	TTE/CTTE  \\	
 6	& bn081003377U &  9:03:06 &	109.504 &	109.728 &	$424 \pm 72$ & $120 \pm 18$	&1, 2 & TTE/CTTE  \\
 7	& bn081003377U &  9:03:06 &	117.488 &	117.568 &	$64 \pm 23 $ & $24 \pm 18 $	&1, 2 & TTE/CTTE  \\	
 8	& bn081003377U &  9:03:06 &	117.904 &	118.048 &	$104 \pm 34$ & $56 \pm 11$	&1, 2 &	TTE/CTTE  \\	
 9	& bn081003377U &  9:03:06 &	149.824 &	149.952 &	$136 \pm 18 $ & $64 \pm 11$	&1, 2 &	TTE/CTTE  \\	
 10	& bn081003377U &  9:03:06 &	240.736 &	240.768 &	\multicolumn{1}{c}{...\tablenotemark{*}} & \multicolumn{1}{c}{...\tablenotemark{*}}	&1, 2 &	TTE/CTTE  \\	
 11	& bn081003377U &  9:03:06 &	306.760 &	\multicolumn{1}{c}{...} &	 \multicolumn{1}{c}{...}  & \multicolumn{1}{c}{...} &	1, 2, 5 &	CTIME  \\
 12	& bn081003377U &  9:03:06 &	321.090 &	\multicolumn{1}{c}{...} &	 \multicolumn{1}{c}{...}  & \multicolumn{1}{c}{...} &	1, 2, 5 &	CTIME  \\
 13	& bn081003377U &  9:03:06 &	411.780 &	\multicolumn{1}{c}{...} &	 \multicolumn{1}{c}{...}  & \multicolumn{1}{c}{...} &	1, 2, 5 &	CTIME   \\
 14	& bn081003377U &  9:03:06 &	468.440 &	\multicolumn{1}{c}{...} &	 \multicolumn{1}{c}{...}  & \multicolumn{1}{c}{...} & 1, 2, 5 &	CTIME   \\
 15	& bn081003377U &  9:03:06 &	584.137 &	585.161 &  \multicolumn{1}{c}{...} &	 \multicolumn{1}{c}{...} &	1, 2, 5 &	CSPEC \\
16	& bn081003385 &  9:14:00 &	-0.168 &	0.008 &	$108 \pm 18 $ & $ 56 \pm 34 $ &	1, 2, 5 &	TTE/CTTE  \\
17	& bn081003385U &  9:14:00 &	103.432 &	103.544 &	\multicolumn{1}{c}{...\tablenotemark{*}} & \multicolumn{1}{c}{...\tablenotemark{*}} &	2, 5 &	TTE/CTTE  \\	
18	& bn081003446 &  10:42:53 &	-0.038 &	0.090 &	$120 \pm 14 $ & $ 60 \pm 11$ &1, 2, 5 &	TTE/CTTE  \\	
19	& bn081003779 &  18:41:39 &	-0.023 &	0.105 &	$104 \pm \;\, 9 $ & $ 60 \pm 11$	 & 1, 2, 5 &	TTE/CTTE  \\	
20  & bn081004050 &  1:11:32 & -0.023 & 0.105 &  $132 \pm 28$ & $56 \pm 11$ & 1, 2, 5 &	TTE/CTTE  \\
21	& bn081005020 &  0:29:09 &	-0.023 &	-0.007 &	$24 \pm 16 $ & $12 \pm\;\, 6$& 6, 7 &	TTE/CTTE  \\	
22	& bn081010537  & 12:53:38 &	-0.039 &	0.009  &	$60 \pm 18$ & $24 \pm\;\, 9$  &1, 2, 5 &	TTE/CTTE  \\	
 23	& bn090322789 &  18:56:23.75 &	-0.087& 0.489 &	$592 \pm 40$ & $360\pm 29$	& 0, 1, 3 &	TTE/CTTE  \\	
 24	& bn090322944 &  22:39:15.75 &	-0.023& 0.489	&   $288 \pm \;\,6 $ & $88 \pm \;\, 6 $ & 7, 8 &	TTE/CTTE  \\
 25 & bn090326625 & 15:00:36.01 & -0.023 & 0.089 & $244 \pm 30$ & $80 \pm\;\, 6$ & 9, 10 &	TTE/CTTE  \\	
 26 & bn090326625U & 15:00:36.01 & 2.176 & 2.496 & $192 \pm 45$ & $96 \pm 36$ &9, 10   &	TTE/CTTE  \\
 27	& bn090328545 &  13:05:12.04 &	-0.023& 0.057 &	$136 \pm 37$ & $48 \pm\;\, 9$	 &9, 10 &	TTE/CTTE  \\	
 28	& bn090329754 &  18:05:15.76 &	-0.023& 0.025 &	$140\pm 54$ & $28 \pm 11$	& 10, 11 &	TTE/CTTE  \\	
 29	& bn090330237 &  05:41:09.99 &	-0.023& 0.057 &	$124 \pm 24 $ & $32 \pm \;\,9$	&  6, 7 & TTE/CTTE  \\	
 30	& bn090401093 &  02:14:28.53 &	-0.375 & -0.327	 & $2656 \pm 172 $ & $1632 \pm 172$ & 3, 6, 7	 & TTE/CTTE  \\	
 31	& bn090401666 &  15:59:36.83 &	-0.023 & 0.89 &	$88\pm \;\, 6$ & $32\pm\;\, 6$	& 10 &	TTE/CTTE  \\	
 32	& bn090403592 &  14:13:04.46 &	-0.023 & 0.121 &	$320 \pm 23$ & $80 \pm 36$	& 9, 10 &	TTE/CTTE  \\	
 33	& bn090403761 &  18:15:36.30 &	-0.039& 0.041 &	$136 \pm 18$ & $48 \pm 23$	& 6, 7 &	TTE/CTTE  \\	
 34	& bn090409351 &  08:25:24.01 &	-0.023& 0.073& $112 \pm 18$ & $72 \pm 18$	& 9, 10 &	TTE/CTTE  \\	
 35	& bn090411917 &  22:01:05.27 &	-0.039& 0.137 &	$180 \pm 20 $ & $64 \pm\;\, 9$	& 7, 8 &	TTE/CTTE  \\	
 36	& bn090413987 &  23:41:48.86 &	-0.007& 0.057 & $72 \pm \;\, 6 $ & $28 \pm \;\,6$	&	6, 7 &	TTE/CTTE  \\	
 37	& bn090417946 &  22:42:11.37 &	-0.023& 0.409 &	$480 \pm 69 $ & $336 \pm 23 $	&	 9, 10 &	TTE/CTTE
\enddata
\tablecomments{Table}
\tablenotetext{*}{rmfit crashed} \tablenotetext{a}{yymmdd.thousandth of day} \tablenotetext{b}{related to $T_0 =$ Trigger time}
\end{deluxetable}


\begin{deluxetable}{crrrrrrrrrrrrr}
\renewcommand{\tabcolsep}{2pt}
\renewcommand{\arraystretch}{1.3}
\tabletypesize{\tiny}
\rotate
\tablecaption{Spectral analysis results  \label{tbl-2}}
\tablewidth{0pt}
\tablehead{
\colhead{Burst} &
\multicolumn{3}{c}{BB} &
\multicolumn{3}{c}{OTTB} &
\multicolumn{4}{c}{Comp} &
\colhead{$\Delta$C-stat} &
\colhead{Fluence} &
\colhead{16 ms Peak Flux} \\
\# &

\colhead{A} & \colhead{kT} & \colhead{C-stat} &
\colhead{A} & \colhead{kT} & \colhead{C-stat} &
\colhead{A} & \colhead{E$_{\rm Peak}$} & \colhead{$\alpha$} & \colhead{C-stat} & \colhead{(BB }
&  \colhead{(8-200 keV)} & \colhead{(8-200 keV)} \\
& $\times 10^{-3}$ & &\colhead{/DOF} & $\times 10^{-3}$ & &\colhead{/DOF} & & \colhead{(keV)} & & \colhead{/DOF}&\colhead{ - OTTB)} &	\colhead{($10^{-8}$erg/cm$^2$)} & \colhead{(ph/cm$^2$/s)}
}
\startdata
1 &$4.7^{+1.8}_{-1.3}$ & $11.5^{+1.0}_{-1.0}$ & 165.3/237 &$16.8^{+4.9}_{-4.0}$   & $44.7^{+9.1}_{-6.6}$ & 169.5/237  &... &... &... &... & $-4.2$ & 4.1 $\pm$ 0.5& 14.6 $\pm$ 3.0\\
2 & $3.1^{+0.5}_{-0.5}$ & $12.5^{+0.6}_{-0.5}$ & 215.3/237 &$17.1^{+2.8}_{-1.9}$ & $52.5^{+7.2}_{-4.2}$ & 275.5/237 &... &... &... &...& $-60.2$ & 13.4$\pm$ 0.9& 23.0 $\pm$ 3.4\\
3 & $1.2^{+0.5}_{-0.3}$ & $11.4^{+1.0}_{-1.0}$ & 264.2/237 &$4.8^{+1.4}_{-1.2}$ & $49.5^{+11.2}_{-8.4}$ & 270.8/237&... &... &... &...& $-6.6$& 8.1$\pm$ 0.9 & 9.9 $\pm$ 2.6\\
4 & $2.5^{+0.8}_{-0.6}$ & $12.8^{+1.0}_{-0.9}$ & 194.1/237 &$17.6^{+3.1}_{-3.1}$ & $59.5^{+10.4}_{-8.8}$ & 210.6/237 &... &... &... &... & $-16.6$& 6.5$\pm$ 0.6& 23.1 $\pm$ 3.3\\
5 & $2.6^{+0.7}_{-0.5}$ & $13.1^{+0.8}_{-0.8}$ & 193.7/237 &$18.4^{+3.1}_{-2.9}$ & $56.0^{+8.4}_{-6.7}$ & 226.1/237 & ... &... &... &... & $-32.4$& 7.6$\pm$ 0.7& 26.3 $\pm$ 3.5\\
6 & $2.3^{+0.5}_{-0.4}$ & $12.3^{+0.7}_{-0.6}$ & 221.2/237 &$12.2^{+2.1}_{-1.7}$ & $52.5^{+7.5}_{-5.4}$ & 256.1/237& ... &... &... &... & $-35.0$& 13.1$\pm$ 1.0& 21.7 $\pm$ 3.4\\
7 & $4.0^{+1.6}_{-1.1}$ & $10.8^{+1.0}_{-0.9}$ & 218.4/237 &$14.8^{+3.9}_{-3.6}$ & $54.5^{+13.4}_{-10.0}$ & 226.7/237 & ... &... &... &... & $-8.3$& 4.5$\pm$0.5& 18.6 $\pm$ 3.2\\
8 & $0.7^{+0.5}_{-0.3}$ & $13.6^{+2.1}_{-1.8}$ & 230.7/237 &$8.0^{+2.4}_{-2.4}$ & $70.3^{+30.5}_{-0.0}$ & 228.4/237 & ... &... &... &... &$+ 2.3$ & 3.6$\pm$ 0.6 & 8.3 $\pm$ 2.4\\
9 & $3.2^{+0.6}_{-0.5}$ & $13.2^{+0.6}_{-0.6}$ & 211.2/237 &$25.2^{+3.4}_{-3.0}$ & $59.0^{+7.5}_{-5.9}$ & 237.3/237 & ... &... &... &... & $-26.1$& 1.3$\pm$ 0.1& 26.5 $\pm$ 3.6\\
10& $4.7^{+2.1}_{-1.5}$ & $11.1^{+1.2}_{-1.0}$ & 159.1/237 &$13.2^{+7.2}_{-3.9}$ & $42.9^{+17.1}_{-0.0}$ & 172.3/237 & ... &... &... &... & $-13.2$ & 2.3$\pm$ 0.4 & 22.6 $\pm$ 3.5\\
1-10 &  $ 2.3^{+0.2}_{-0.2}$ & $12.3^{+0.3}_{-0.3}$ &  237.5/237 & $12.8^{+0.8}_{-0.8}$ & $54.2^{+2.9}_{-2.5}$ & 402.7/237 & $10.9^{+5.6}_{-5.6}$ & 47.7$^{+1.1}_{-1.1}$ &1.51$^{+0.27}_{-0.26}$ & 238.3/236& $$-165.2$$ & ... & ...\\
15 & $5.1^{+0.2}_{-0.2}$ & $12.0^{+1.2}_{-1.1}$& 356.7/359& $3.0^{+0.6}_{-0.6}$ & $56.2^{+11.3}_{-9.3}$ & 354.5/359 &$0.19^{+0.33}_{-0.17}$ &$49.9^{+12.6}_{-5.7}$ &$0.25^{+0.47}_{-1.11}$ & 352.7/358 & $+2.2$ & 11.2$\pm$ 0.9 & ... \\
16 & $1.2^{+0.4}_{-0.3}$ & $12.5^{+1.0}_{-0.9}$& 338.6/340& $6.3^{+1.8}_{-1.3}$ &$50.5^{+11.8}_{-6.7}$ &  359.9/340 &$86.5^{+2160}_{-53.0}$ &$47.4^{+3.2}_{-3.2}$ &$2.91^{+1.84}_{-0.96}$ & 336.9/339 & -21.3 & 5.4$\pm$ 0.6  &	 18.0 $\pm$ 2.9 \\
17 & $4.4^{+0.8}_{-0.7}$ & $12.6^{+0.6}_{-0.6}$& 187.6/228& $26.2^{+3.3}_{-3.5}$ & $55.7^{+6.0}_{-5.8}$ & 218.2/228 &$6.3^{+32.8}_{-3.3}$ &$49.5^{+3.9}_{-3.4}$ &$0.84^{+0.97}_{-0.35}$ & 187.5/227 & -30.6 &12.7$\pm$ 0.6   &	 28.6 $\pm$ 2.8 \\
18 & $5.4^{+0.9}_{-0.7}$ & $12.5^{+0.5}_{-0.5}$& 329.4/347& $30.1^{+3.1}_{-3.1}$& $51.0^{+4.1}_{-4.0}$& 348.5/347 &$4.1^{+4.8}_{-2.1}$ & $49.4^{+2.4}_{-2.2}$& $0.53^{+0.40}_{-0.36}$ & 324.8/346  & $-19.1$& 17.3$\pm$ 0.7  &	 39.1 $\pm$ 3.0 \\
19 & $8.8^{+0.1}_{-0.1}$ & $10.7^{+0.4}_{-0.4}$& 383.7/358& $22.8^{+2.7}_{-2.7}$& $42.9^{+3.3}_{-3.1}$& 392.8/358 &$3.12^{+3.03}_{-1.45}$ & $42.4^{+2.0}_{-1.8}$&$0.26^{+0.33}_{-0.31}$ &371.5/357 & $-9.1$& 15.0$\pm$ 0.5  &	 44.8 $\pm$ 2.7  \\
20 &  $4.0^{+0.6}_{-0.5}$ & $12.6^{+0.5}_{-0.5}$ & 327.1/360 & $24.9^{+2.7}_{-2.6} $ & $55.7^{+5.3}_{-4.7}$& 369.3/360 & $9.6^{+13.3}_{-5.2}$	 &49.2$^{+2.3}_{-2.1}$ & 1.14$^{+0.44}_{-0.40}$ &326.5/359 & $-42.2$ & 13.0$\pm$ 0.5& 51.7 $\pm$ 4.0\\
21 &  $1.1^{+0.4}_{-0.3}$ & $11.9^{+1.0}_{-1.0}$ & 143.1/241 & $52.1^{+11.8}_{-11.2}$& $49.3^{+8.9}_{-7.4}$& 154.6/241 & $4.9^{+15.4}_{-3.3}$& $47.9^{+5.3}_{-4.4}$& $0.3^{+0.7}_{-0.6}$& 140.4/240 & $-11.5$& 3.8$\pm$ 0.3   &	 43.05 $\pm$ 3.1 \\
22 & $2.6^{+0.7}_{-0.6}$ & $14.0^{+1.0}_{-0.9}$ & 305.3/358 & $26.4^{+4.2}_{-4.1}$ & $61.5^{+10.1}_{-8.4}$ & 328.5/358 & $20.4^{+105.0}_{-14.5}$ &$53.9^{+3.9}_{-3.5}$ &$1.7^{+1.0}_{-0.7}$ & 305.4/357& $-23.2$&  5.0$\pm$ 0.3  &	 20.35 $\pm$ 1.9 \\
23 & $3.5^{+0.9}_{-0.6}$ & $10.1^{+0.5}_{-0.6}$& 562.8/351 & $9.5^{+1.1}_{-1.1}$ & $47.9^{+4.2}_{-3.8}$& 442.7/351 & $0.02^{+0.01}_{-0.01}$ & $43.3^{+8.7}_{-8.1}$ & $-1.63^{+0.18}_{-0.16}$ & 431.31/350& $+120.1$& 30.9 $\pm$ 1.6  &	 34.3.0 $\pm$ 2.7 \\
25 & $12.3^{+3.2}_{-2.5}$ & $9.4^{+0.6}_{-0.6}$ &268.8/231 & $ 18.7^{+3.8}_{-3.5}$ & $ 40.0^{+4.8}_{-4.2}$ &  232.7/231 & $183^{+177}_{-83}$ & 39.7$^{+5.1}_{-4.9}$ & -1.1$^{+1.1}_{-0.8}$ & 232.6/230 & $+36.1$ &  13.2$\pm$ 0.9 &  55.0 $\pm$ 5.9\\
26 &  $0.5^{+0.3}_{-0.2}$ & $13.8^{+2.0}_{-1.9}$& 299.2/254 & $7.6^{+1.5}_{-1.6}$ & $105.0^{+48.8}_{-30.0}$& 298.2/254 & $0.10 \pm 0.14 $ & $ 69.7 \pm 15.8 $ & $ -0.18 \pm 0.78$ & 297.11/253 & +1.0 & $ 9.5 \pm 3.3$ & $ 7.7 \pm 1.1$\\
27 & $16.7^{+3.8}_{-3.0}$ & $9.4^{+0.5}_{-0.5}$& 202.9/235& $23.8^{+4.7}_{-4.3}$  & $39.0^{+4.3}_{-3.8}$   & 177.6/235& $0.64^{+0.73}_{-0.32}$& $39.2^{+3.6}_{-3.2}$& $-0.66^{+0.35}_{-0.32}$& 176.4/234 & $+25.3$& 12.6$\pm$ 0.6  &	 53.3 $\pm$ 4.2 \\
28 & $29.3^{+6.9}_{-5.2}$ & $9.4^{+0.5}_{-0.5}$& 191.9/238& $35.2^{+7.3}_{-6.6}$  & $35.6^{+3.8}_{-3.4}$   & 173.1/238& $1.47^{+2.0}_{-0.8}$&$36.8^{+3.2}_{-3.1}$ &$-0.56^{+0.40}_{-0.36}$ & 171.6/237 & $+18.8$& 12.7$\pm$ 0.6  &	 97.7 $\pm$ 6.3  \\
29 & $97.4^{+8.5}_{-7.2}$ & $8.9^{+0.2}_{-0.2}$& 417.9/239 & $79.7^{+6.8}_{-6.5}$  & $32.6^{+1.3}_{-1.2}$   & 226.2/239& $2.23^{+0.77}_{-0.56}$&$33.3^{+1.3}_{-1.4}$ &$-0.87^{+1.4}_{-1.3}$ & 225.3/238 & $+191.7$& 55.7$\pm$ 1.0  &	518 $\pm$ 12 \\
30 & $3.0^{+2.4}_{-1.3}$& $10.7^{+1.7}_{-1.6}$& 278.0/360& $10.3^{+4.4}_{-4.0}$  & $49.4^{+19.7}_{-13.1}$ & 271.2/362 & $0.01 \pm 0.05 $ &$ 76.7 \pm 114.0  $ &$ -1.84 \pm 0.43 $ & 270.0/359& +6.8 & 2.5$\pm$0.3   &	 14.1 $\pm$ 1.7\\
31 &  $261^{+14}_{-13}$ & $8.8^{+0.1}_{-0.1}$& 361.3/118& $176^{+11}_{-11}$& $30.9^{+0.8}_{-0.8}$& 158.8/118& $17.9^{+4.7}_{-3.7}$& $ 33.2^{+0.7}_{-0.7}$ & $ -0.36^{+0.11}_{-0.10}$&  117.0/117 & $+202.5$& 186.7$\pm$ 2.4  &	 1122 $\pm$ 26 \\
32 & $24.3^{+4.3}_{-3.4}$ & $8.5^{+0.4}_{-0.4}$& 309.3/237& $18.4^{+3.0}_{-2.8}$  & $34.1^{+2.7}_{-2.4}$   & 247.5/237&$0.37^{+0.26}_{-0.15}$ &$34.2^{+2.7}_{-2.7}$ &$-0.96^{+0.24}_{-0.23}$ & 247.5/236 & $+61.8$& 21.4 +/ 0.7  &	 92.6 $\pm$ 5.1 \\
33 & $9.8^{+3.6}_{-2.4}$ & $9.5^{+0.8}_{-0.8}$& 246.9/241 & $19.1^{+4.4}_{-4.0}$  & $44.7^{+7.2}_{-5.9}$   & 218.1/241& $0.08^{+0.09}_{-0.04}$& $45.6^{+10.8}_{-8.6}$& $-1.38^{+0.35}_{-0.0}$& 217.0/240 & $+28.8$& 8.4$\pm$ 0.5  &	 42.5 $\pm$ 3.5 \\
34 & $54.0^{+8.1}_{-6.8}$ & $7.9^{+0.3}_{-0.3}$& 251.0/237& $22.0^{+3.9}_{-3.5}$  & $29.0^{+2.0}_{-1.8}$   & 216.1/237&$1.54^{+2.33}_{-0.63}$ & $30.5^{+2.1}_{-2.1}$& $-0.63^{+0.40}_{-0.25}$&  213.7/236 & $+34.9$& 23.3$\pm$ 0.7  &	 149 $\pm$ 7 \\
35 & $41.1^{+4.8}_{-4.0}$ & $8.5^{+0.2}_{-0.2}$& 332.9/241 & $28.2^{+3.1}_{-2.9}$  & $31.8^{+1.5}_{-1.5}$   & 217.5/241& $0.97^{+0.45}_{-0.30}$&$32.6^{+1.5}_{-1.6}$ &$-0.82^{+0.17}_{-0.17}$ & 216.2/240 & $+115.4$& 45.5$\pm$ 1.0  &	 258 $\pm$ 9 \\
36 &  $60.2^{+5.8}_{-5.1}$& $9.7^{+0.2}_{-0.2}$& 295.0/240 & $77.6^{+7.4}_{-7.0}$  & $35.4^{+1.7}_{-1.6}$   & 213.4/240& $3.7^{+1.8}_{-1.1}$& $37.1^{+1.4}_{-1.4}$& $-0.50^{+0.18}_{-0.17}$& 204.6/239 & $+81.6$&37.7$\pm$ 0.8  &	 364 $\pm$ 10 \\
37 &  $5.3^{+1.6}_{-1.1}$ &$9.3^{+0.6}_{-0.6}$ & 331.0/237 & $9.5^{+1.7}_{-1.5}$   & $44.5^{+5.4}_{-4.6}$   & 280.6/237 & $0.03^{+0.02}_{-0.01}$& $45.2^{+8.8}_{-7.8}$& $-1.49^{+0.25}_{-0.0}$& 277.2/236 & $+51.0$ & 22.7$\pm$ 1.0  &	 64.7 $\pm$ 4.6
\enddata
\end{deluxetable}

\end{document}